\documentstyle[aps,pre,twocolumn,epsf,floats]{revtex}

\begin{document}

\date{\today}
\date{September 2000}
\title{Localization-induced coherent backscattering effect in wave dynamics}
\author{H. Schomerus, K. J. H. van Bemmel, and C. W. J. Beenakker}
\address{Instituut-Lorentz, Universiteit Leiden, P.O. Box 9506, 2300 RA
Leiden, The Netherlands}

\twocolumn[
\widetext
\begin{@twocolumnfalse}

\maketitle

\begin{abstract}
We investigate the statistics
of single-mode delay times of waves reflected from a disordered
wave\-guide in the presence of wave localization.
The distribution of delay times is 
qualitatively different from the distribution in the diffusive regime,
and sensitive to coherent backscattering:
The probability to find small delay times is enhanced by a factor close
to $\sqrt{2}$ for reflection angles near the angle of incidence.
This dynamic effect of coherent backscattering disappears
in the diffusive regime.
\end{abstract}

\vspace{1cm}
\end{@twocolumnfalse}
]

\narrowtext

\section{Introduction}
\label{sec1}

The two most prominent interference effects arising from multiple scattering are
coherent backscattering and wave localization
\cite{ishimaru,sheng,berk,lw1,lw2,John}.
Both effects are related to the {\em static} intensity of a wave reflected or transmitted
by a medium with randomly located scatterers. 
Coherent backscattering is the enhancement of the reflected intensity in a
narrow cone around the angle of incidence, and is a result of the systematic
constructive
interference in the presence of time-reversal symmetry \cite{lw1,lw2}.
Localization arises from systematic destructive interference and suppresses
the transmitted intensity \cite{John}.

This paper presents a detailed theory of a recently discovered \cite{letter}
interplay between coherent backscattering and localization in a {\em dynamic}
scattering property,
the single-mode delay time of a wave reflected by a disordered waveguide.
The single-mode delay time 
is the derivative $\phi'={\rm d}\phi/{\rm d}\omega$ of the phase $\phi$
of the wave amplitude with respect to the frequency $\omega$. 
It is linearly related to the Wigner-Smith delay times of scattering
theory \cite{wigner,smith,fs}  and
is the key observable of recent experiments on multiple scattering of
microwaves \cite{vantiggelen:1999a} and light waves \cite{ld}.
Van Tiggelen, Sebbah, Stoytchev, and Genack
\cite{vantiggelen:1999b}
have developed a statistical theory for the distribution of
$\phi'$ in a waveguide geometry (where angles
of incidence are discretized as modes). Although the theory was worked
out mainly for the case of transmission, the implications for reflection
are that the distribution $P(\phi')$ does not depend on whether the
detected mode $n$ is the same as the incident mode $m$ or not.
Hence it appears that no coherent backscattering effect exists
for $P(\phi')$. 

What we will demonstrate here is that this is true only if  
wave localization may be disregarded. The previous studies 
\cite{vantiggelen:1999a,vantiggelen:1999b} dealt with the diffusive
regime of wave\-guide lengths $L$ below the localization length $\xi$.
(The localization length in a waveguide geometry is $\xi\simeq N l$, with $N$ 
the number of propagating modes and $l$ the mean free path.)
Here we consider the localized regime
$L>\xi$ (assuming that also the absorption length $\xi_{\rm a}>\xi$).
The distribution of reflected intensity is insensitive to the presence
or absence of localization, being given in both regimes by Rayleigh's
law.
In contrast, we find that the delay-time distribution changes markedly
as one enters the localized regime, decaying more slowly for large
$|\phi'|$. Moreover, a coherent backscattering effect appears: For $L >
\xi$ the peak of $P(\phi')$ is higher for $n=m$ than for $n\neq m$ by a
factor which is close to $\sqrt{2}$, the precise factor being
$\sqrt{2} \times \frac{4096}{1371\pi}=1.35$.

We also consider what happens if time-reversal symmetry is broken,
by some magneto-optical effect.
The coherent backscattering effect disappears. However,
even for $n\neq m$, the delay-time distribution for preserved time-reversal
symmetry is different than for broken time-reversal symmetry.
This difference is again only present for $L>\xi$ and vanishes in the
diffusive regime.

The plan of this paper is as follows:
In Section \ref{sec2} we specify the notion \cite{vantiggelen:1999a}
of the single-mode delay
time $\phi'$, relate it to the Wigner-Smith delay times
and 
review the results \cite{vantiggelen:1999b}
for the diffusive regime, extending them to include ballistic corrections.
This Section also
contains the random-matrix formulation for the localized regime,
that provides the basis for our calculations,
and includes a brief discussion of the 
conventional coherent backscattering effect in the static intensity
$I$.
Section \ref{sec:dyn}
presents the calculation of the
joint distribution of $\phi'$ and
$I$.
We compare our analytical theory with numerical simulations and 
give a qualitative argument for the dynamic coherent backscattering effect.
The role of absorption is discussed, as well as the effect of
broken time-reversal symmetry.
Details of the calculation are delegated to the appendices.
\section{Delay times}
\label{sec2}
\subsection{Single-mode delay times}

\begin{figure}
\epsfxsize8cm
\center{\epsfbox{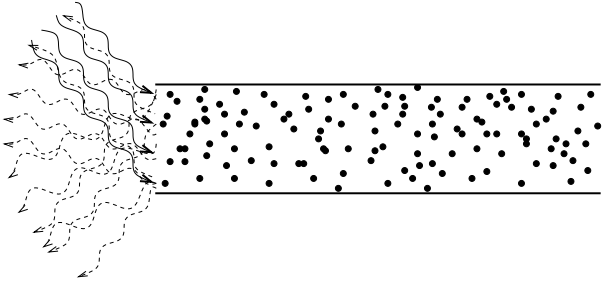}}
\medskip
\medskip
\caption{Sketch of a waveguide containing a randomly scattering medium and
illuminated by a monochromatic plane wave.
We study the frequency dependence
of the phase $\phi$ of the reflected wave amplitude in a single speckle,
corresponding to a single waveguide mode. The derivative
$\phi'={\rm d}\phi/{\rm d}\omega$ is the single-mode delay time.
}
\label{fig:geometry}
\end{figure}

We consider a disordered medium (mean free path $l$) in a waveguide geometry
(length $L$), as depicted in Fig.\ \ref{fig:geometry}.
There are $N\gg 1$ propagating modes at frequency $\omega$,
given by
$N=\pi{\cal A}/{\lambda^2}$   
for a waveguide with an opening of area ${\cal A}$.
The wave velocity is $c$, and we consider a scalar wave (disregarding
polarization) for simplicity. In the numerical simulations we will work
with a two-dimensional waveguide of width $W$, where $N=2 W/\lambda$.

We study the dependence of the reflected wave amplitude
\begin{equation}
r_{nm}=\sqrt{I}e^{i\phi}
\end{equation}
on the frequency $\omega$.
The indices $n$ and $m$ specify the detected and incident mode,
respectively.
(We assume single-mode excitation and detection.)
Here $I=|r_{nm}|^2$ is the intensity of the reflected wave
in the detected mode for unit incident intensity, and characterizes the
static properties of the reflected wave.
Dynamic information is contained in the 
phase derivative
\begin{equation}
\phi'=\frac{{\rm d}\phi}{{\rm d} \omega},
\end{equation}
which has the dimension of a time and is called the single-mode delay time
\cite{vantiggelen:1999a,vantiggelen:1999b}.
The intensity $I$ and the delay time $\phi'$
can be recovered from the product of reflection matrix elements
\begin{equation}
\rho=r_{nm}(\omega+\case{1}{2}\delta\omega)r_{nm}^*(\omega-\case{1}{2}\delta\omega)
\;,
\label{eq:corr}
\end{equation}
evaluated at two nearby frequencies $\omega\pm\frac{1}{2}\delta\omega$.
To leading order in the frequency difference $\delta\omega$ one has
\begin{equation}
\rho=I(1+i\delta\omega\phi')
\Rightarrow
I=\lim_{\delta \omega\to 0}\mbox{Re}\,\rho,\,\,
\phi'=\lim_{\delta \omega\to 0}
\frac{{\rm Im}\,\rho}{\delta\omega \, I}
\;.
\label{eq:corriphi}
\end{equation}

We seek the joint distribution function $P(I,\phi')$ in an ensemble of different
realizations of disorder. We distinguish between the
diffusive regime where $L$ is small compared to the localization
length $\xi\simeq Nl$, and the localized regime where $L\gtrsim \xi$.
Localization also requires that the absorption length $\xi_{\rm a}\gtrsim\xi$.
We will contrast the case of excitation and detection in two distinct
modes $n\neq m$ with the equal-mode case $n = m$.
Although we mainly focus on the optically more relevant case of
preserved time-reversal symmetry, we will also discuss the case
of broken time-reversal symmetry for comparison.
These two cases are indicated by the index $\beta=1$, 2, respectively.

\subsection{Relation to Wigner-Smith delay times}

In the localized regime ($\xi\ll L,\xi_{\rm a}$) we can relate
the single-mode delay time  $\phi'$ to the 
Wigner-Smith \cite{wigner,smith,fs} delay times $\tau_i$,
with $i=1,\ldots,N$.
The $\tau_i$'s
are defined for a unitary reflection matrix $r$ (composed of the
elements $r_{nm}$), hence they require the absence of transmission and
of absorption.
One then has
\begin{mathletters}
\begin{eqnarray}
&&-i r^\dagger\frac{{\rm d}r}{{\rm d}\omega}=
U^\dagger\,{\rm diag}\,(\tau_1,\ldots,\tau_{N}) U,\\
&&-i r^*\frac{{\rm d}r^T}{{\rm d}\omega}=
V^\dagger\,{\rm diag}\,(\tau_1,\ldots,\tau_{N}) V,
\end{eqnarray}%
\end{mathletters}
with $U$ and $V$ unitary matrices of eigenvectors.
In the presence of time-reversal symmetry $r$ is a symmetric matrix,
hence $V=U$ in this case.

For small $\delta\omega$ we can expand 
\begin{equation}
r(\omega\pm\case{1}{2}\delta \omega)= V^{\rm T} U \pm
\case{1}{2}i\,\delta\omega\,
 V^{\rm T}\,{\rm diag}\,(\tau_1,\ldots,\tau_N)\, U 
\;.
\label{eq:poldec}
\end{equation}
Inserting it into Eq.\ (\ref{eq:corr}) and comparison with
Eq.\ (\ref{eq:corriphi}) yields the relations
\begin{equation}
\label{eq:corr2}
\phi'={\rm Re}\,\frac{A_1}{A_0}\;,\quad I=|A_0|^2\;,\quad A_k=
\sum_i\tau_i^k u_i v_i\;.
\end{equation}
We have abbreviated $u_i=U_{im}$, $v_i=V_{in}$.
In the special case $n=m$ 
the coefficients $u_i$ and $v_i$ 
are identical in the presence of time-reversal symmetry.

The distribution of the Wigner-Smith delay times in the
localized regime was
determined recently \cite{BB}. In terms of the rates $\mu_i=1/\tau_i$
it has the form of the Laguerre ensemble of random-matrix theory,
\begin{equation}
P(\{\mu_i\})\propto \prod_{i<j}|\mu_i-\mu_j|^\beta\prod_{k}\Theta(\mu_k)
e^{-\gamma (\beta N +2-\beta)\mu_k}
\;,
\label{eq:laguerre}
\end{equation}
where the step function
$\Theta(x)=1$ for $x>0$ and $0$ for $x<0$.
The parameter $\gamma$ is defined by
\begin{equation}
\gamma=\alpha l/c,
\label{eq:gamma}
\end{equation}
with the coefficient $\alpha=\pi^2/4$ or $8/3$ for two- or
three-dimensional scattering, respectively.
Eq.\ (\ref{eq:laguerre}) extends the $N=1$ result 
of Refs.\ \cite{Jayannavar,Heinrichs,Comtet} to any $N$.

The matrices $U$ and $V$ in Eq.\ (\ref{eq:poldec})
are uniformly distributed in the unitary group.
They are independent for $\beta=2$, while $U=V$ for $\beta=1$.
In the large-$N$ limit the matrix elements become
independent
Gaussian random numbers with vanishing mean and variance $1/N$.
Hence
\begin{equation}
\langle u_i\rangle =\langle v_i \rangle =0,\quad
\langle |u_i|^2\rangle =\langle |v_i|^2\rangle =N^{-1},
\label{eq:gau}
\end{equation}
with $u_i=v_i$ for $n=m$ and $\beta=1$.
Corrections to this Gaussian approximation are of order $1/N$.

\subsection{Diffusion theory}

The joint probability distribution $P(I,\phi')$ in the diffusive regime
$l\ll L\ll\xi$
has been derived in Refs.\ \cite{vantiggelen:1999a,vantiggelen:1999b},
\begin{eqnarray}
&&P_{\rm diff}(I,\phi')=
\Theta(I)(I/\pi{\bar I}^3)^{1/2}e^{- I/\bar I}
\nonumber
\\
&&{}\quad{}\times
(Q \bar{\phi'}^2)^{-1/2}
\exp\left(-\frac{I}{\bar I}\frac{(\phi'-\bar{\phi'})^2}{Q\bar{\phi'}^2}\right)
\;,
\label{eq:iphifixed}
\end{eqnarray}
with constants ${\bar I}$, $\bar{\phi'}$, and $Q$.
It has the same form for transmission and
reflection, the only difference being the dependence of the constants on
the system parameters.
Here we focus on the case of reflection, because we are concerned with
coherent backscattering.

From the joint distribution function (\ref{eq:iphifixed}) one obtains
for the intensity the Rayleigh distribution
\begin{equation}
P_{\rm diff}(I)=\frac{1}{\bar I}\exp(-I/{\bar I}).
\label{eq:rayleigh}
\end{equation}
Hence $\bar I$ is the mean detected intensity per mode.
It is given by \cite{MelloStone}
\begin{equation}
\bar I=\frac{1}{N} (1+\delta_{\beta 1}\delta_{nm}),
\label{eq:imean0}
\end{equation}
assuming unit incident intensity.
The factor-of-two enhancement in the case $n=m$
is the static coherent
backscattering effect mentioned in the introduction, which
exists only in the presence of time-reversal symmetry ($\beta=1$).
Eqs.\ (\ref{eq:rayleigh}) and (\ref{eq:imean0}) remain valid in the
localized regime, since they are determined by scattering on the scale
of the mean free path. Hence $L\gg l$ is sufficient for static coherent
backscattering, and it does not matter whether $L$ is small or large
compared to $\xi$.

By integrating over $I$ in Eq.\ (\ref{eq:iphifixed}) one arrives at 
the distribution of single-mode delay times
\cite{vantiggelen:1999a,vantiggelen:1999b},
\begin{equation}
P_{\rm diff}(\phi')=
\frac{Q}{2\bar{\phi'}}[Q+(\phi'/\bar{\phi'}-1)^2]^{-3/2}
\;.
\label{eq:pphidiff}
\end{equation}
Hence $\bar{\phi'}$ is the mean delay time while 
$\sqrt{Q}$ sets the relative width of the distribution.
These constants are determined by the correlator
\cite{vantiggelen:1999a,vantiggelen:1999b}
\begin{eqnarray}
C_{12}&=&\frac{\langle r_{nm}(\omega+\delta\omega)r_{nm}^*(\omega)
\rangle}{\langle r_{nm}(\omega)r_{nm}^*(\omega)\rangle}
\nonumber\\
&=&1+i{\bar{\phi'}}\,\delta\omega-\case{1}{2}{\bar{\phi'}}^2(Q+1)(\delta\omega)^2.
\end{eqnarray}
Diffusion theory gives
\begin{equation}
 \bar{\phi'}= 2\gamma s/3,\quad Q= 2s/5.
\end{equation}
Here $\gamma$ is given by Eq.\ (\ref{eq:gamma}). We have
defined
\begin{equation}
s=\alpha' L/l,
\label{eq:s}
\end{equation}
where the numerical coefficient
$\alpha'= 2/\pi$, 3/4 for two-, three-dimensional scattering.
(The corresponding result for $Q$ given in Ref.\ \cite{vantiggelen:1999b} 
is incorrect.)

Diffusion theory predicts 
that the distribution of delay times (\ref{eq:pphidiff}) as well as
the values of the constants $\bar{\phi'}$ and $Q$
do not depend on the choice $n=m$ or $n\neq m$ (and also not on whether
time-reversal symmetry is preserved or not).
Hence there is no dynamic effect of coherent backscattering in the
diffusive regime.

\subsection{Ballistic corrections}

The expressions for the constants ${\bar I}$, ${\bar{\phi'}}$, and $Q$
given above are valid up to corrections of order $l/L$. 
Here we give
more accurate formulas that account for these ballistic corrections.
(We need these to compare with numerical simulations.)
We determine the ballistic corrections for $Q$ and $\bar{\phi'}$ 
by relating the dynamic problem
to a static problem with absorption.
(This relationship only works for the mean. It cannot be used to
obtain the distribution \cite{BeenBem}.)
The mean total reflectivity
\begin{equation}
\bar a
=
1+x-\sqrt{2x+x^2}\coth\left[s\sqrt{2x+x^2}+\,\mbox{arcosh}\,(1+x)\right]
\label{eq:bpb}
\end{equation} 
for absorption $\alpha' x$ per mean free path
has been evaluated in Ref.\ \cite{bpb}.
[Here $\alpha'$ is the same constant as in the definition of $s$, Eq.\
(\ref{eq:s}).]
We identify $C_{12}={\bar a(x)}/{\bar a(0)}$
by analytic continuation to an imaginary
absorption rate $x=-i\delta\omega\gamma$.
Expanding in $x$ to second order we find
\begin{equation}
\label{eq:qp}
\bar{\phi'}=\gamma\frac{s(3+2s)}{3(1+s)}
,\quad
Q=\frac{8s^3+28s^2+30s+15}{5(2s+3)^2}
.
\end{equation}

\subsection{Numerical simulation}
\label{sec:num}

\begin{figure}
\epsfxsize8cm
\center{\epsfbox{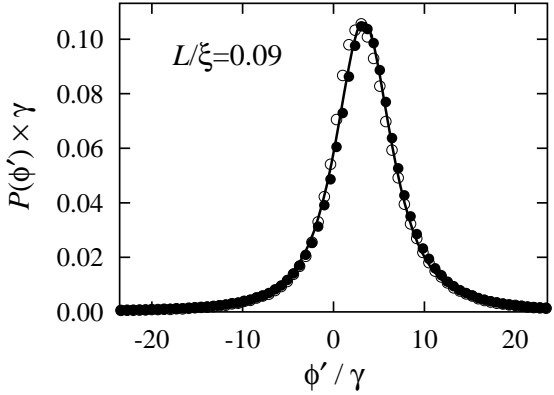}}
\medskip
\caption{Distribution of the single-mode delay time $\phi'$ 
in the diffusive regime.
The result of numerical simulation (data points)
with $N=50$ propagating modes
is compared to the prediction  (\ref{eq:pphidiff})
of diffusion theory (solid curve).
There is no difference between the case $n=m$ of equal-mode excitation
and detection (open circles) and the case $n\neq m$ of excitation and
detection in distinct modes (full circles).
}
\label{fig1}
\end{figure}

The validity of diffusion theory
was tested in Refs.\ \cite{vantiggelen:1999a,ld,vantiggelen:1999b}
by comparison with experiments in transmission.
In Fig.\ \ref{fig1} we show an alternative test in reflection,
by comparison with a numerical simulation of scattering
of a scalar wave by a two-dimensional random medium.
(We assume time-reversal symmetry.)
The reflection matrices
$r(\omega\pm\frac{1}{2}\delta\omega)$ are computed by applying the method of
recursive Green functions \cite{recgf} to the Helmholtz equation
on a square lattice (lattice constant $a$). The width $W=100\,a$ and the
frequency $\omega=1.4\,c/a$ are chosen such that there are $N=50$
propagating modes. The mean free path
$l=14.0\,a$ is found from
the formula \cite{beenrev} $\mbox{tr}\,rr^\dagger=Ns(1+s)^{-1}$
for the reflection probability.
The corresponding localization length $\xi=NL/s=1100\,a$.
The parameter $\gamma=46.3\,a/c$ is found
from Eq.\ (\ref{eq:qp}) by
equating $\langle\phi'\rangle=\bar{\phi'}$.
(This value of $\gamma$ is somewhat bigger than
the value $\gamma= \pi^2 l/4c=34.5\,
a/c$ expected for two-dimensional scattering,
as a consequence of the anisotropic
dispersion relation on a square lattice.)
We will use the same set of parameters in the interpretation of the
results in the localized regime, later in this paper.

Our numerical results confirm that in the diffusive regime the
distribution of delay times $\phi'$ does not distinguish between 
excitation and detection in distinct modes
($n\neq m$, full circles) and identical modes
($n=m$, open circles).

\section{Dynamic coherent backscattering effect}
\label{sec:dyn}
\subsection{Distinct-mode excitation and detection}
\label{sec5}

We now calculate the joint probability distribution
function $P(I,\phi')$ of intensity $I$ and single-mode delay
time $\phi'$ in the localized regime,
for the typical case $n\neq m$ of excitation and detection
in two distinct modes.
We assume preserved time-reversal symmetry ($\beta=1$), leaving the case 
of broken time-reversal symmetry for the end of this section.

It is convenient to work momentarily with
the weighted delay time
$W=\phi' I$ and to recover $P(I,\phi')$ from $P(I,W)$ at the end.
The characteristic function
\begin{equation}
\chi(p,q)=\left\langle e^{-ip I-iq W}\right\rangle
\end{equation}
is the Fourier transform of $P(I,W)$.
The average 
$\langle \cdots\rangle$
is over the vectors ${\bf u}$ and ${\bf v}$ and over the set
of eigenvalues $\{\tau_i\}$. The average over one of the vectors, say
${\bf v}$, is easily carried out, because it
is a Gaussian integration. The result is a determinant, 
\begin{mathletters}
\begin{eqnarray}
&&\chi(p,q)=\left\langle\det(1+i H/N)^{-1}\right\rangle
\label{eq:chi1}
\;,
\\
&&H=p{\bf u}^*{\bf u}^{\rm T}+\case{1}{2}q (\bar{\bf u}^*
{\bf u}^{\rm T}+{\bf u}^*\bar{\bf u}^{\rm T})\;.
\end{eqnarray}
\end{mathletters}
The Hermitian matrix $H$ is a sum
of dyadic products of the vectors ${\bf u}$ and $\bar{\bf u}$,
with  $\bar{u}_i=u_i \tau_i$,
and hence has only two non-vanishing eigenvalues $\lambda_+$ and
$\lambda_-$. Some straightforward linear algebra gives
\begin{equation}
\lambda_\pm=\case{1}{2}\left(q B_1+p\pm\sqrt{2pq B_1+q^2B_2+p^2}\right)
\;,
\end{equation}
where we have defined the spectral moments
\begin{equation}
B_k=\sum_i|u_i|^2\tau_i^k
\;.
\label{eq:ab}
\end{equation}
The resulting determinant is 
\begin{equation}
\det(1+ H/N)^{-1}=(1+\lambda_+/N)^{-1}
(1+\lambda_-/N)^{-1},
\end{equation}
hence
\begin{equation}
\chi(p,q)
=\left\langle
\left[1+\frac{ip}{N}+\frac{iq}{N}B_1+\frac{q^2}{4N^2}(B_2-B_1^2)\right]^{-1}
\right\rangle \;.
\end{equation}
An inverse Fourier transform, followed by a change of variables from
$I$, $W$ to $I$, $\phi'$, gives
\begin{eqnarray}
&&P(I,\phi')=
\Theta(I)(N^3 I/\pi)^{1/2}e^{-N I}
\nonumber
\\
&&{}\quad{}\times
\left\langle
(B_2-B_1^2)^{-1/2}
\exp\left(-NI\frac{(\phi'-B_1)^2}{B_2-B_1^2}\right)
\right\rangle
.
\label{eq:iphi}
\end{eqnarray}
The average is over the spectral moments
$B_1$ and $B_2$, which depend on the
$u_i$'s and $\tau_i$'s via Eq.\ (\ref{eq:ab}).

The calculation 
of the joint distribution $P(B_1,B_2)$
is presented in Appendix \ref{app:pab}. The result is
\begin{eqnarray}
&&P(B_1,B_2)=\Theta(B_1)\Theta(B_2)\exp\left(-\frac{N B_1^2}{B_2}\right)
\nonumber\\ 
&&{}\times\left[
\frac{B_1^2\gamma N^3}{B_2^4}(B_2+\gamma N^2 B_1)
\exp\left(-\frac{2\gamma N}{B_1}\right)
\right.
\nonumber\\
&&\left.{}
-\frac{\gamma^3 N^5}{4B_2^5}(2B_2^2-4B_1^2B_2N+B_1^4N^2)
{\rm Ei}\,\left(-\frac{2\gamma N}{B_1}\right)
\right]
,
\label{eq:pab1}
\end{eqnarray}
where ${\rm Ei}\,(x)$ is the 
exponential-integral function.
The distribution $P(I,\phi')$ follows from 
Eq.\ (\ref{eq:iphi})
by integrating over $B_1$
and $B_2$ with weight given by
Eq.\ (\ref{eq:pab1}). 

\begin{figure}
\epsfxsize8cm
\center{\epsfbox{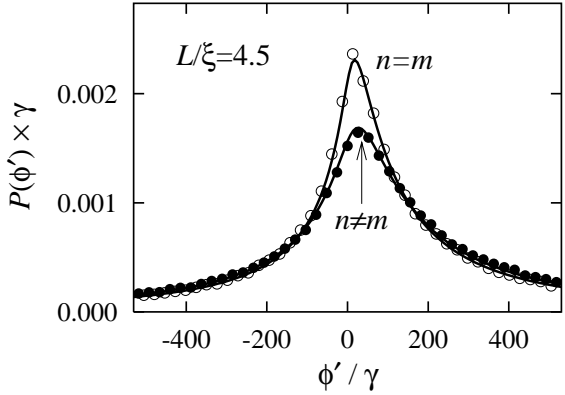}}
\medskip
\caption{Distribution of the single-mode delay time $\phi'$ 
in the localized regime.
The results of numerical simulations
with $N=50$ propagating modes
(open circles for $n=m$, full circles for
$n\neq m$)
are compared to the analytical predictions.
The curve for different incident and detected modes $n\neq m$
is obtained from Eqs.\ (\ref{eq:pab1}) and (\ref{eq:pphi}).
The curve for $n=m$ is calculated from
Eqs.\ (\ref{eq:phicd}) and (\ref{eq:pcdnn}).
The same value for $\gamma$ is used as in the diffusive regime,
Fig.\ \ref{fig1}.
}
\label{fig2}
\end{figure}

Irrespective of the distribution of $B_1$ and $B_2$, we recover  from
Eq.\ (\ref{eq:iphi}) the Rayleigh law (\ref{eq:rayleigh}) for the
intensity $I$.
The distribution
$P(\phi')=\int_0^\infty{\rm d}I\,P(I,\phi')$
of the single-mode delay time takes the form
\begin{equation}
P(\phi')=\int\limits_0^\infty{\rm d}B_1
\int\limits_0^\infty {\rm d}B_2\,
\frac{P(B_1,B_2)(B_2-B_1^2)}{2(B_2+\phi'^2-2B_1\phi')^{3/2}}
\;.
\label{eq:pphi}
\end{equation}
In Fig.\ \ref{fig2} this distribution
is compared with the result of a numerical simulation 
of a random medium as in
Section \ref{sec:num}, but now in the localized regime.
The same value for $\gamma$ was used as in Fig.\ \ref{fig1},
making this comparison a parameter-free test of the theory.
(Note that $\gamma$ alone determines the complete distribution
function in the localized regime, 
in contrast to the diffusive case where two parameters 
are required.)
The numerical data agrees very well with the analytical prediction.

\subsection{Equal-mode excitation and detection}
\label{sec6}

We now turn to the  case $n = m$ of equal-mode excitation and detection,
still assuming that time-reversal symmetry is preserved.
Since $u_i=v_i$, we now have
\begin{equation}
\label{eq:phicd}
\phi'={\rm Re}\,\frac{C_1}{C_0}
\;,\quad I=|C_0|^2\;,\quad
C_k=\sum_i \tau_i^k u_i^2
\;.
\end{equation}
The joint distribution function $P(C_0,C_1)$
of these complex numbers
can be calculated in the same way
as $P(B_1,B_2)$. In Appendix \ref{app:a0a1} we obtain
\begin{eqnarray}
P(C_0,C_1)\propto
\exp(-N |C_0|^2/2)
\int_0^\infty{\rm d} s\,
s^2 e^{-s} &&
\nonumber \\
{}\times
\left(1+\frac{|C_1|^2 s^2}{\gamma^2 N^2}-
\frac{2s}{\gamma N} \,{\rm Re}\, C_0 C_1^*\right)^{-5/2}
\;.&&
\label{eq:pcdnn}
\end{eqnarray}
The corresponding distribution function 
$P(\phi')$ is plotted also in Fig.\ \ref{fig2}
and compared with the results of the numerical simulation.
Good agreement is obtained, without any free parameter.

\subsection{Comparison of both situations}
\label{sec6a}

Comparing the two curves in Fig.\ \ref{fig2},
we find a striking difference
between distinct-mode and equal-mode excitation
and detection:
The distribution for $n=m$ displays an enhanced
probability of small delay times. 
In the vicinity of the peak, both distributions become very similar
when the delay times for $n\neq m$ are divided by a scale factor of
about $\sqrt{2}$. 
In limit $N\to\infty$ (see following sub-section), the maximal value 
$P(\phi'_{\rm peak})=\sqrt{2/\pi N^3\gamma^2}$ for $n=m$
is larger than the maximum of $P(\phi')$ for $n\neq m$
by a factor 
\begin{equation}
\frac{P_{n=m}(\phi'_{\rm peak})}{P_{n\neq m}(\phi'_{\rm peak})}
=\sqrt{2} \times \frac{4096}{1371\pi}=1.35
\,.
\label{eq:fac}
\end{equation}
Correspondingly, the probability to find very large delay times is reduced
for $n=m$. This is reflected by the asymptotic behavior
\begin{equation}
P(\phi')\sim
\frac{\gamma N^{3/2}}{\phi'^2}\times
\left\{\begin{array}{cc}
(2\pi)^{-1/2}&\qquad\mbox{for }n=m,\\
\sqrt{\pi}/4&\qquad\mbox{for }n\neq m.
\end{array}
\right.
\label{eq:phitail}
 \end{equation}

The enhanced probability of small delay times for $n=m$
is the dynamic coherent backscattering effect mentioned in the introduction.
The effect requires localization and is
not observed in the diffusive regime.

\subsection{Limit $N\to\infty$}
\label{sec:rescor}

\begin{figure}
\epsfxsize8cm
\center{\epsfbox{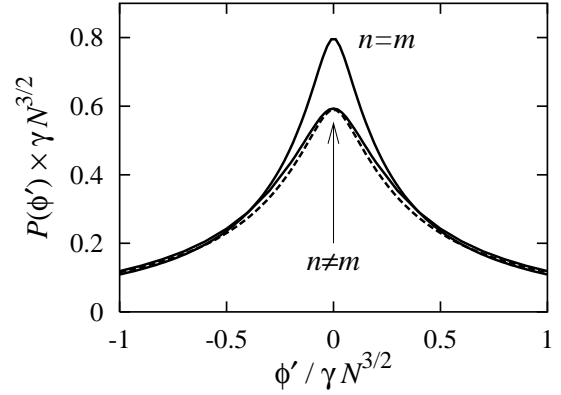}}
\medskip
\caption{Distribution of the single-mode delay time $\phi'$ 
in the localized regime for preserved time-reversal symmetry,
in the limit $N\to \infty$. In this limit
$P(\phi')$ becomes symmetric
for positive and negative values of $\phi'$.
Compared are the result for $n\neq m$ [Eqs.\ (\ref{eq:pphisym}),
(\ref{eq:pbbeta1})]
and $n=m$  [Eqs.\ (\ref{eq:phicd}), (\ref{eq:pat0}), (\ref{eq:pcnn})].
The distribution for $n=m$ falls on top of the distribution
for $n\neq m$ when $\phi'$ is rescaled by a factor 1.35 (dashed curve,
almost indistinguishable from the solid curve for $n\neq m$).
}
\label{fig:pphisym}
\end{figure}

The results presented so far assume $N\gg 1$, but retain finite-$N$
corrections
of order $N^{-1/2}$. (Only terms of order $1/N$ and higher are
neglected.)
It turns out that the asymmetry
of $P(\phi')$ for positive and negative
values of $\phi'$ is an effect of order $N^{-1/2}$.
The asymmetry is hence captured faithfully by our calculation.
We now consider how the asymmetry
eventually disappears in the limit $N\to\infty$.

For distinct modes $n \neq m$,  the spectral moments scale as
$B_1\sim \gamma N$ and $B_2\sim \gamma^2 N^3$.
With
$\phi'\sim \gamma N^{3/2}$, one finds that $B_1$ can be omitted
to order $N^{-1/2}$ in Eq.\ (\ref{eq:pphi}). One obtains
the symmetric distribution
\begin{equation}
P(\phi')=\int_0^\infty{\rm d}B_2\,
\frac{P(B_2)B_2}{2(B_2+\phi'^2)^{3/2}}
,
\label{eq:pphisym}
\end{equation}
plotted in Fig.\ \ref{fig:pphisym}.

For identical modes $n=m$,
observe that the quantities 
$C_0$ and $C_1$ become
mutually independent in the large-$N$ limit:
The cross-term $(\gamma N)^{-1} \,{\rm Re}\, C_0 C_1^*$ 
in Eq.\ (\ref{eq:pcdnn})
is of relative order
$N^{-1/2}$ because $C_0\sim N^{-1/2}$ and $C_1\sim \gamma N$.
Hence to order $N^{-1/2}$
the distribution factorizes,
$P(C_0,C_1)=P(C_0)P(C_1)$.
The distribution of $C_0$ is a Gaussian,
\begin{equation}
P(C_0)=\frac{N}{2\pi }\exp(-N|C_0|^2/2),
\label{eq:pat0}
\end{equation}
as a consequence of the central-limit theorem, and
\begin{equation}
P(C_1)\propto
\int_0^\infty{\rm d} s\,
s^2 e^{-s} 
\left(1+\frac{|C_1|^2 s^2}{\gamma^2 N^2}\right)^{-5/2}
\;.
\label{eq:pcnn}
\end{equation}
The resulting 
distribution of $\phi'={\rm Re}\,(C_1/C_0)$
is also plotted in Fig.\ \ref{fig:pphisym}.

The dynamic coherent backscattering effect persists in
the limit $N\to\infty$, it is therefore not due
to finite-$N$ corrections.
The peak heights differ by the factor given in Eq.\ (\ref{eq:fac}).

\subsection{Interpretation in terms of large fluctuations}
\label{sec:qual}

In order to explain
the coherent backscattering enhancement of the peak of $P(\phi')$
in more qualitative terms,
we compare Eq.\ (\ref{eq:phicd}) for $n=m$ with the
corresponding relation (\ref{eq:corr2}) for $n\neq m$.

The factorization of the joint distribution function
$P(C_0,C_1)$ discussed in the previous sub-section
can be seen as a consequence of
the high density of anomalously large Wigner-Smith
delay times $\tau_i$
in the Laguerre ensemble (\ref{eq:laguerre}).
The distribution of the largest time $\tau_{\rm max}=\max_i\tau_i$
follows from the distribution of the smallest eigenvalue in the
Laguerre ensemble, calculated by Edelman \cite{edelman}.
It is given by
\begin{equation}
P(\tau_{\rm max})=\frac{\gamma N^2}{\tau_{\rm max}^2}
\exp(-\gamma N^2/\tau_{\rm max}).
\end{equation}
As a consequence, the spectral moment $C_1$ is dominated by a small 
number of contributions $u_i^2\tau_i$
(often enough by a single one, say with index $i=1$), while $C_0$ can be safely
approximated by the sum over all remaining indices $i$ (say, $i\neq 1$).
The same argument applies also to the spectral moments $A_k$
which determine the delay-time
statistics for $n\neq m$, hence the distribution function $P(A_0,A_1)$
factorizes as well.

The quantities $A_0$ and $C_0$ have a Gaussian distribution for
large $N$, because of the central-limit theorem,
with $P(C_0)$ given by Eq.\ (\ref{eq:pat0}) and
\begin{equation}
P(A_0)=\frac{N}{\pi }\exp(-N|A_0|^2).
\label{eq:pa0}
\end{equation}
It becomes then clear that
the main contribution to the enhancement (\ref{eq:fac}) of the peak height,
namely the factor of
$\sqrt{2}$,
has the same origin as the 
factor-of-two enhancement of the mean intensity $\bar I$.
More precisely, the relation $P(A_0=x)=2\,P(C_0=\sqrt{2}\,x)$
leads to a rescaling of $P(I)$ for $n=m$ by a factor of $1/2$
and to a rescaling of $P(\phi')$ by a factor of
$\sqrt{2}$.
The remaining factor of
$\frac{4096}{1371\pi}= 0.95$ comes from the difference in the
distributions $P(A_1)$ and $P(C_1)$.
It turns out that the distribution
\begin{eqnarray}
&&P(A_1)
=\int_0^\infty{\rm d}s\, \frac{1}{4\pi\gamma N} \frac{s^2}{(4+|A_1/\gamma
N|^2s^2)^3}
\nonumber\\
&&{}\times \left[e^{-s}(64 +32 s+12 s^2+s^3)-3s^2{\rm Ei}\,(-s)\right]
\label{eq:pa1t}
\end{eqnarray}
(derived in Appendix \ref{appa1})
is very similar to $P(C_1)$ given in Eq.\ (\ref{eq:pcnn}),
hence the remaining factor is close to unity.

The large $\tau_i$'s  are related to the penetration of the 
wave deep into the localized regions and are
eliminated in the diffusive regime $L\lesssim\xi$.
In the following sub-section we compare the localized 
and diffusive regimes in more detail.

\subsection{Localized versus diffusive regime}
\label{sec5a}

\begin{figure}
\epsfxsize8cm
\center{\epsfbox{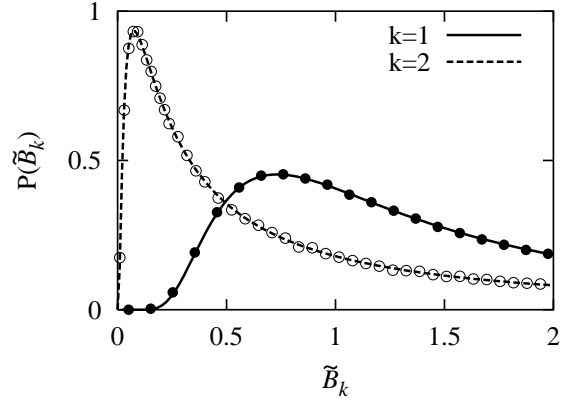}}
\caption{Distributions of $\tilde B_1=B_1/\gamma N$ and
$\tilde B_2=B_2/\gamma^2 N^3$.
The analytic prediction from Eq.\ (\ref{eq:pab1})
[for explicit formulas see Eqs.\ (\ref{eq:pa}) and (\ref{eq:pbbeta1})]
is compared to the result of a numerical simulation of the
Laguerre ensemble with $N=50$.}
\label{fig:pb}
\end{figure}

Comparison of Eqs.\ (\ref{eq:iphifixed}) and (\ref{eq:iphi}) shows that
the two joint distributions
of $I$ and $\phi'$ 
would be identical if statistical fluctuations in the
spectral moments $B_1$, $B_2$ could be ignored. 
The correspondences are 
\begin{equation}
B_1\leftrightarrow \bar{\phi'},\quad B_2-B_1^2\leftrightarrow Q \bar{\phi'}^2
.
\end{equation}
However, 
the distribution $P(B_1,B_2)$ is very broad (see Fig.\ \ref{fig:pb}), so that
fluctuations can {\em not} be ignored.
The most probable values are
\begin{equation}
B_1^{\rm typical}\simeq \gamma N,\quad
B_2^{\rm typical}\simeq\gamma^2 N^3,
\end{equation}
but the mean values $\langle B_1\rangle$, $\langle
B_2\rangle$ diverge---demonstrating the presence of large fluctuations.
In the diffusive regime $L\lesssim\xi$ the spectral moments
$B_1$ and $B_2$ can be replaced by their ensemble averages,
and the diffusion theory \cite{vantiggelen:1999a,vantiggelen:1999b}
is recovered. (The same applies if
the absorption length $\xi_{\rm a}\lesssim \xi$.)

The large fluctuations in $B_1$ and $B_2$ directly affect
the statistical properties of the delay time $\phi'$.
We compare the distribution (\ref{eq:pphi})
in the localized regime  (Fig.\ \ref{fig2}) with the
result (\ref{eq:pphidiff}) of diffusion theory (Fig.\ \ref{fig1}).
In the localized regime the value
$\phi'_{\rm peak}\simeq B_1^{\rm typical}$
at the center of the peak of $P(\phi')$
is much smaller than the width of the peak
$\Delta\phi'\simeq  (B_2^{\rm typical})^{1/2}\simeq \phi'_{\rm
peak}(\xi/l)^{1/2}$. This holds also in the diffusive regime, where 
$\phi'_{\rm peak}=\bar{\phi'}$ and
$\Delta\phi'\simeq \phi'_{\rm peak}(L/l)^{1/2}$.
However, the mean $\langle \phi'\rangle=\langle B_1\rangle$ diverges for $P$,
but is finite (equal to $\bar{\phi'}$) for $P_{\rm diff}$.
For large $B_2$ one has asymptotically
$P(B_2)\sim \frac{1}{4} N\gamma^{3/2}\sqrt{\pi} B_2^{-3/2}$. As a consequence,
in the tails $P(\phi')$ 
falls off only quadratically [see Eq.\ (\ref{eq:phitail})],
while in the diffusive regime
$P_{\rm diff}(\phi')\sim \frac{1}{2}Q\bar{\phi'}^2|\phi'|^{-3}$
falls off with an inverse third power.

\subsection{Role of absorption}
\label{sec:abs}

Although absorption causes the same exponential decay of the
transmitted intensity as localization, this decay is of a quite different,
namely incoherent nature.
The strong fluctuations in the localized regime disappear
as soon as the absorption length $\xi_{\rm a}$ drops below the localization
length $\xi$, because long paths which penetrate into the localized
regions are suppressed by absorption.
In this situation one should expect that the results of
diffusion theory are again valid even for $L\gtrsim\xi$.
This expectation is confirmed by our numerical
simulations.
(We do not know how to incorporate absorption effects into our analytical
theory.)

\begin{figure}
\epsfxsize8cm
\center{\epsfbox{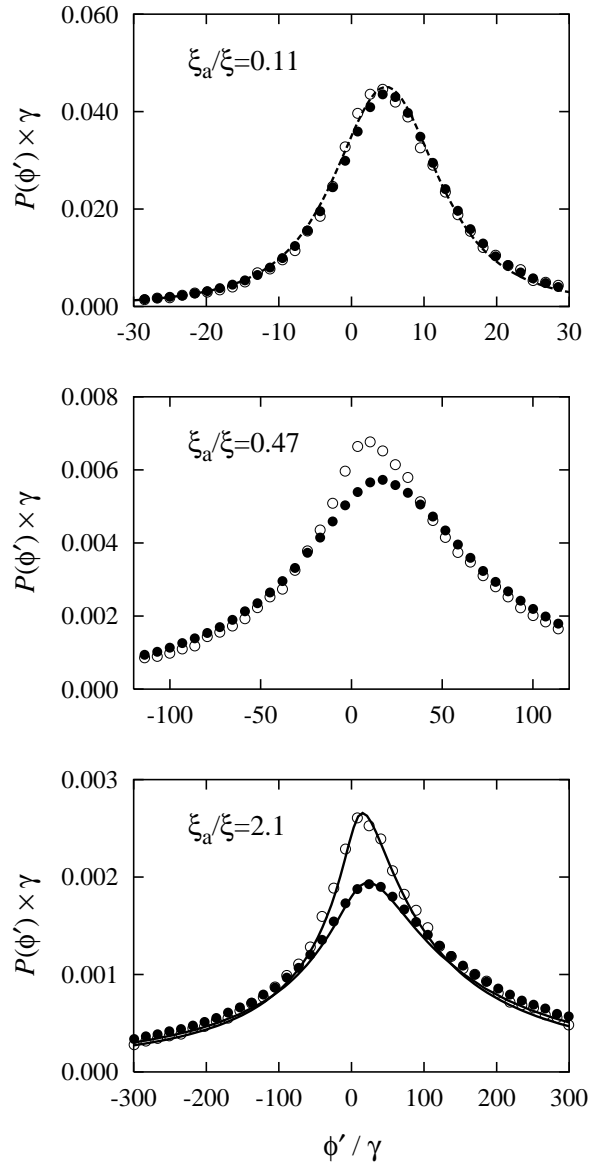}}
\caption{Single-mode delay-time distribution $P(\phi')$ in the presence of
absorption. The data points are the result of a numerical simulation
of a waveguide with length $L=4.5\xi$.
Open circles are for equal-mode excitation and detection $n=m$,
full circles for the case of distinct modes $n\neq m$.
In the upper panel 
(with  $\xi_{\rm a}<\xi$), the
data is compared to the prediction (\ref{eq:pphidiff})
of diffusion theory.
In the lower panel we compare with the predictions 
(\ref{eq:pab1}-\ref{eq:pcdnn})
of random-matrix theory.}
\label{fig:abs}
\end{figure}

In Fig.\ \ref{fig:abs} we plot the delay-time distribution
for two values of the
absorption length $\xi_{\rm a}<\xi$ and one value
$\xi_{\rm a}>\xi$, both for equal-mode and for
distinct-mode excitation and detection.
The length of the waveguide is $L=4.1\,\xi$.
The result for strong absorption with 
$\xi_{\rm a}=0.11\xi$ is very similar 
to Fig.\ \ref{fig1}. Irrespective of the choice of the detection mode,
the data can be fitted to the prediction (\ref{eq:pphidiff}) of diffusion
theory.
The plot for $\xi_{\rm a}=0.47\xi$ shows that
the dynamic coherent backscattering effect slowly sets in
when the absorption length becomes comparable
to the localization length.
The data also deviates from the 
prediction of diffusion theory.
The full factor (\ref{eq:fac}) between the peak heights
quickly develops as soon as $\xi_{\rm a}$ exceeds $\xi$,
as can be seen from the data for $\xi_{\rm a}=2.1\xi$.
Moreover, this data can already be fitted to the predictions of
random-matrix theory, with 
$\gamma\approx 53.2\,a/c$. 
(The value $\gamma=46.3\,a/c$ of Sec.\ \ref{sec:num}
is reached when absorption is further reduced.)

\subsection{Broken time-reversal symmetry}
\label{sec7}

\begin{figure}
\epsfxsize8cm
\center{\epsfbox{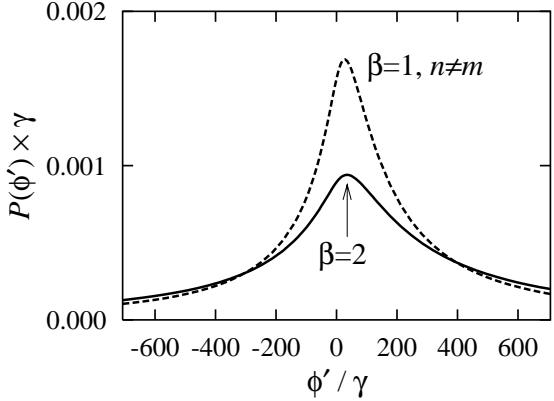}}
\medskip
\caption{
Comparison of the single-mode delay-time distributions for preserved
and broken time-reversal symmetry.
The number of propagating modes is $N=50$.
The curves are calculated from Eq.\ (\ref{eq:pphi}) 
with $P(B_1,B_2)$ given by Eq.\ (\ref{eq:pab1}) ($\beta=1$)
or Eq.\ (\ref{eq:pbb2}) ($\beta=2$).}
\label{fig4}
\end{figure}

The case $\beta=2$
of broken time-reversal symmetry is less important for optical
applications, but has been realized in microwave experiments
\cite{erbacher:93,alt:95,stoffregen:95}.
There is now no difference between $n=m$ and $n\neq m$.
The matrices $U$ and $V$ have the same statistical
distribution as for the case of preserved time-reversal symmetry. Hence,
by following the steps of Section \ref{sec5}, we arrive again at
Eq.\ (\ref{eq:iphi}), with spectral moments $B_k$ as defined in Eq.\
(\ref{eq:ab}). Their joint distribution has now to be calculated from 
Eq.\ (\ref{eq:laguerre}) with $\beta=2$. This calculation
is carried out in Appendix \ref{app:2}. The result is
\begin{equation}
P(B_1,B_2)=\frac{2\gamma N^3B_1^2}{B_2^3}\exp(-N B_1^2/B_2-2\gamma N/B_1)
.
\label{eq:pbb2}
\end{equation}
The distribution of single-mode delay times $P(\phi')$ is 
given by Eq.\ (\ref{eq:pphi}) with this new function $P(B_1,B_2)$.
We plot $P(\phi')$
in Fig.\ \ref{fig4} and compare it to the case of preserved
time-reversal symmetry. The distribution 
is rescaled by about a factor of
two towards larger delay times when time-reversal symmetry is broken.
This can be understood from the fact
that the relevant length scale, the localization length, 
is twice as big for broken time-reversal
symmetry ($\xi=2NL/s$, while $\xi=NL/s$ for preserved time-reversal symmetry). 

\section{Conclusion}

We have presented a detailed theory, supported by 
numerical simulations, of a recently discovered
\cite{letter} coherent backscattering effect
in the single-mode delay times of a wave reflected by a disordered waveguide.
This dynamic effect is special because it
requires localization for its existence,
in contrast to the static coherent backscattering effect in the reflected
intensity.
The dynamic effect can be understood from the combination of the
static effect and the large fluctuations in the localized regime.

In the diffusive regime there is no dynamic coherent backscattering effect:
The distribution of delay times is unaffected by the choice of the
detection mode and the presence or absence of time-reversal symmetry.
The effect also disappears when the absorption length is smaller than the
localization length. In both situations the large fluctuations 
characteristic for the localized regime are suppressed.

Existing experiments on the delay-time distribution
\cite{vantiggelen:1999a,ld}
have verified the diffusion theory \cite{vantiggelen:1999b}.
The theory for the localized regime presented here awaits
experimental verification.

\begin{acknowledgments}
We thank P. W. Brouwer for valuable advice. This work was 
supported by the ``Nederlandse organisatie voor Wetenschappelijk Onderzoek''
(NWO) and by the ``Stichting voor Fundamenteel Onderzoek der Materie'' (FOM).
\end{acknowledgments}

\appendix

\section{Joint distribution of $B_1$ and $B_2$ for $\beta=1$}

\label{app:pab}

We calculate the joint probability distribution function
$P(B_1,B_2)$ of the spectral moments $B_1$ and $B_2$,
defined in Eq.\ (\ref{eq:ab}), which
determine $P(I,\phi')$ from Eq.\ (\ref{eq:iphi}).
We assume preserved time-reversal symmetry ($\beta=1$).
Since $B_k=\sum_i|u_i|^2\mu_i^{-k}$,
we have to average over the wavefunction amplitudes
$u_i$, which are Gaussian complex numbers with zero mean and
variance $1/N$,
and the rates $\mu_i$ which are distributed according to the Laguerre
ensemble (\ref{eq:laguerre}) with $\beta=1$.
This Laguerre ensemble is represented as the eigenvalues of an
$N\times N$ Hermitian matrix $W^\dagger W$,
where $W$ is a complex symmetric matrix with the Gaussian distribution
\begin{equation}
P(W)\propto\exp[-\gamma(N+1){\rm tr}\,W^\dagger W].
\label{eq:pw1}
\end{equation}
The calculation is performed neglecting corrections of order $1/N$,
so that we are
allowed to replace $N+1$ by $N$.
The measure is
\begin{equation}
{\rm d}W=\prod_{i<j}{\rm d}\,{\rm Re}\,W_{ij}\,{\rm d}\,{\rm Im}\,W_{ij}\,
\prod_{i}{\rm d}\,{\rm Re}\,W_{ii}\,{\rm d}\,{\rm Im}\,W_{ii}
\,.
\end{equation}

\subsection{Characteristic function}

In the first step we express $P(B_1,B_2)$ by its characteristic function,
\begin{eqnarray}
P(B_1,B_2)&=&\frac{1}{(2\pi)^2}
\int_{-\infty}^\infty{\rm d}p\,\int_{-\infty}^\infty
{\rm d}q\,e^{ipB_1+iqB_2}\chi(p,q)
\;,
\nonumber\\
\label{eq:pchi}\\
\chi(p,q)&=&
\left\langle
\prod_{l=1}^N\exp\left[-i|u_l|^2\left(\frac{p}{\mu_l}+\frac{q}{\mu_l^2}
\right)\right]
\right\rangle,
\end{eqnarray}
and average over the $u_l$'s,
\begin{eqnarray}
\chi(p,q)&=&
\left\langle
\prod_{l=1}^N\left(1+i\frac{p}{\mu_l N}+i \frac{q}{\mu_l^2 N}\right)^{-1}
\right\rangle
\nonumber\\
&=&
\left\langle\frac{{\rm det}\,(W^\dagger W)^2}{{\rm det}\,[(W^\dagger
W)^2+ip(W^\dagger W)/N+iq/N]}
\right\rangle
\;.
\nonumber\\
\end{eqnarray}
We have expressed the product over eigenvalues as a ratio of
determinants.
We write the determinant in the denominator as an
integral over a complex vector ${\bf z}$,
\begin{eqnarray}
&&\chi(p,q)\propto
\int {\rm d}W \int {\rm d}{\bf z}
\exp[-\gamma N{\rm tr}\,W^\dagger W]{\rm det}\,(W^\dagger W)^2
\nonumber\\
&&{}\times
\exp\{-{\bf z}^\dagger[(W^\dagger W)^2+ip(W^\dagger W)/N+iq/N]{\bf z}\}
\;.
\label{eq:a5}
\end{eqnarray}
This integral converges because $W^\dagger W$ is positive definite.

\subsection{Parameterization of the matrix $W$}

Now we choose a parameterization of $W$ which facilitates a stepwise
integration over its degrees of freedom.
The distribution of $W$ is
invariant under transformations $W\to U^T W U$, with any unitary matrix $U$.
Hence we can choose  a basis in which ${\bf z}$ points
into direction 1 and write W in block form
\begin{equation}
W=\left(\begin{array}{cc}
a & {\bf x}^T \\
{\bf x} & X
\end{array}
\right).
\end{equation}
Here $a$ is a complex number.
For any $N-1$ dimensional vector ${\bf x}$
we can use another unitary transformation
on the $X$ block
after which ${\bf x}$ points into direction 2.
Then $W$ is of the form
\begin{equation}
W=\left(\begin{array}{ccc}
a & x &0^T \\
x & b & {\bf y}^T \\
0 & {\bf y} & Y
\end{array}
\right),
\label{eq:parw}
\end{equation}
with the real number $x=|{\bf x}|$.
In this parameterization
\begin{eqnarray*}
&&(W^\dagger W)_{11}=
|a|^2+x^2
\;,
\\{}
&&[(W^\dagger W)^2]_{11}=
(|a|^2+x^2)^2+x^2y^2+x^2|a+b^*|^2
\;,
\\
&&{\rm det}\,W=[a(b-{\bf y}^TY^{-1}{\bf y})-x^2]{\rm det}\,Y
\;,
\\
&&{\rm tr}\,W^\dagger W=|a|^2+|b|^2+2x^2+2y^2+{\rm tr}\,Y
\;,\\
&&{\rm d}W={\rm d}^2a\,{\rm d}^2b\,{\rm d}{\bf x}\,{\rm d}{\bf y}\,{\rm d}Y
\;,
\end{eqnarray*}
with $y=|{\bf y}|$.
A suitable transformation on $Y$
allows to replace the term ${\bf y}^T Y^{-1}{\bf
y}$ by $y^2(Y^{-1})_{11}$.

For this parameterization of $W$,
the integrand in Eq.\ (\ref{eq:a5}) depends on the vectors ${\bf x}$, ${\bf
y}$,
and ${\bf z}$ only by their magnitudes $x$, $y$, and $z=|{\bf z}|$. 
Hence we can replace ${\rm d}{\bf x}\to x^{2N-3}\,{\rm d}x$,
 ${\rm d}{\bf y}\to y^{2N-5}\,{\rm d}y$, and
 ${\rm d}{\bf z}\to z^{2N-1}\,{\rm d}z$.

\subsection{Integration}

The integrand in Eq.\ (\ref{eq:a5})
involves $z$, $p$, and $q$ in the form
\begin{equation}
\exp[-z^2([(W^\dagger W)^2]_{11}+ip (W^\dagger W)_{11}/N+iq/N)]
\;.
\end{equation}
It is convenient to pass back to $P(B_1,B_2)$ by Eq.\ (\ref{eq:pchi}), 
because 
integration over $p$ and $q$ gives delta functions
\begin{eqnarray}
&&\delta[B_1-z^2(|a|^2+x^2)/N]\delta(B_2-z^2/N)
\nonumber\\
&&=
\frac{1}{B_2}\delta(B_1/B_2-|a|^2-x^2)\delta(B_2-z^2/N)\;.
\label{eq:a14}
\end{eqnarray}
Subsequent integration over $z$
results in
\begin{eqnarray}
&&P(B_1,B_2)\propto
\int {\rm d}^2a\,{\rm d}^2b\,{\rm d}{x}\,{\rm d}{y}\,x^{2N-3}y^{2N-5}
\nonumber\\
&&{}\times
B_2^{N-2}\delta(B_1/B_2-|a|^2-x^2)
\nonumber\\
&&{}\times
[c_0|ab-x^2|^4+4c_2|a|^2y^4 |ab-x^2|^2+c_4|a|^4y^8]
\nonumber\\
&&{}\times
\exp\left[
-NB_2\left(\frac{B_1^2}{B_2^2}+x^2 y^2+x^2|a^*+b|^2\right)
-2\gamma N y^2
\right]
.
\nonumber\\
\label{eq:b17}
\end{eqnarray}
Here we omitted a term
$\gamma N(|a|^2+|b|^2+2 x^2)$ in the exponent, because
it is of order $1/N$ as we shall see later. 
Furthermore, we denoted
\begin{equation}
c_m=\frac{\langle |{\rm det}\,Y|^4 |(Y^{-1})_{11}|^m\rangle}{
\langle |{\rm det}\,Y|^4\rangle}\;.
\label{eq:b18}
\end{equation}
These coefficients will be calculated later, with the result 
$c_0=1$, 
$c_2=2\gamma$, $c_4=4\gamma^2$.
Integration over $y$ yields for the terms proportional to $c_m$ the factors
$(B_2  x^2+2\gamma )^{-m-N+2}$,
which can be combined with the factor
$(B_2 x^2)^{N-2}$, giving to order $1/N$
[we anticipate $\gamma/B_2 x^2={\cal O}(1/N)$]
\begin{equation}
\frac{
(B_2 x^2)^{N-2}}{(B_2 x^2+2\gamma)^{N-2+m}}
\to(B_2 x^2)^{-m}\exp\left(-\frac{2\gamma N}{B_2 x^2}\right)
.
\label{eq:expo}
\end{equation}

We introduce a new integration variable by $b'=b+a^*$.
So far $P(B_1,B_2)$ is reduced to the form
\begin{eqnarray}
&&P(B_1,B_2)\propto
\int {\rm d}^2a\,{\rm d}^2b'\,{\rm d}x\, x
\delta(B_1/B_2-|a|^2-x^2)
\nonumber\\
&&{}\times
\left(\left|ab'-\frac{B_1}{B_2}\right|^4
+\frac{4c_2|a|^2}{B_2^2x^4}
\left|ab'-\frac{B_1}{B_2}\right|^2
+\frac{c_4|a|^4}{B_2^4x^8}\right)
\nonumber\\
&&{}\times
\exp\left(-\frac{2\gamma N}{x^2 B_2}-N\frac{B_1^2}{B_2}
-NB_2x^2|b'|^2
\right)
\;.
\label{eq:B20}
\end{eqnarray}
Let us now convince ourselves
with this expression that we were justified to omit 
the term $\gamma N(|a|^2+|b|^2+2x^2)$ in Eq.\ (\ref{eq:b17})
and to use Eq.\ (\ref{eq:expo}).
Indeed, 
the various quantities scale as
$B_1\simeq\gamma N$, 
$B_2\simeq\gamma^2 N^3$, and  $|a|^2\simeq |b|^2\simeq x^2 \simeq
1/\gamma N^2$, because any
$\gamma$ and $N$ dependence
disappears if one passes to appropriately rescaled quantities
$B_1/\gamma N$, etc.
The terms omitted are therefore of order $1/N$.

The remaining integrations in Eq.\ (\ref{eq:B20}) are readily performed,
with the final result (\ref{eq:pab1}).
The distribution of $B_1$ to order $1/N$ is 
\begin{equation}
P(B_1)=\frac{\gamma N}{B_1^3}(B_1+2\gamma N)
\exp\left(-\frac{2\gamma N}{B_1}\right)\;.
\label{eq:pa}
\end{equation}
The spectral moment $B_1$ appeared before in a different physical context
in Ref.\ \cite{bpb}, but only a heuristic approximation was given in
that paper.  Eq.\ (\ref{eq:pa}) solves this random-matrix problem precisely.

For completeness we also give the
distribution of the other spectral moment $B_2$
(rescaled as
${\tilde B}_2=B_2 \gamma^{-2} N^{-3}$)
in terms of Meijer $G$ functions,
\begin{eqnarray}
&&P({\tilde B}_2)=\frac{1}{64{\tilde B}_2^{5/2}\pi^{1/2}}\bigg[
14\pi {\tilde B}_2
-16\, G^{0,3}_{3,0}({\tilde B}_2|-\case{1}{2},0,\case{3}{2})
\nonumber\\ &&{}
+20\, G^{0,3}_{3,0}({\tilde B}_2|-\case{1}{2},\case{1}{2},1)
+22\, G^{0,3}_{3,0}({\tilde B}_2|\case{1}{2},\case{1}{2},1)
\nonumber\\ &&{}
+8\, G^{0,3}_{3,0}({\tilde B}_2|\case{1}{2},1,\case{3}{2})
+4\, G^{0,3}_{3,0}({\tilde B}_2|\case{1}{2},\case{3}{2},2)
\nonumber\\
&&{}
-8\,G^{1,3}_{4,1}\left({\tilde B}_2\left|\begin{array}{c}
-\case{1}{2},0,\case{3}{2},2\\1\end{array}
\right.\right)
-16\,G^{1,3}_{4,1}\left({\tilde B}_2\left|\begin{array}{c}
0,\case{1}{2},\case{3}{2},2\\1\end{array}
\right.\right)
\nonumber\\ &&{}
+3\,G^{0,4}_{4,1}\left({\tilde B}_2\left|\begin{array}{c}
\case{1}{2},\case{1}{2},\case{3}{2},\case{3}{2}\\0\end{array}
\right.\right)
\bigg]
\;.
\label{eq:pbbeta1}
\end{eqnarray}

\subsection{Coefficients}

Now we calculate the coefficients $c_2$ and $c_4$ defined in Eq.\
(\ref{eq:b18}).
It is convenient to resize the matrix $Y$ to dimension $N$ (instead of $N-2$)
and to set momentarily $\gamma N=1$.
We use again a block decomposition,
\begin{equation}
Y=\left(\begin{array}{cc}
a&{\bf w}^T\\
{\bf w}&Z
\end{array}
\right)
\;,
\end{equation}
and employ the identities
\begin{mathletters}
\begin{eqnarray}
{\rm det}\,Y&=&(a-{\bf w}^T Z^{-1}{\bf w}){\rm det}\,Z\;,
\\
(Y^{-1})_{11}&=&(a-{\bf w}^T Z^{-1}{\bf w})^{-1}
\;.
\end{eqnarray}
\end{mathletters}
Hence 
\begin{equation}
c_4=\frac{\langle |{\rm det}\,Z|^4\rangle}{\langle |{\rm
det}\,Y|^4\rangle}=\frac{4}{(N+1)(N+3)}\;,
\end{equation}
where we used Selberg's integral \cite{Mehta} for
\begin{equation}
\langle|{\rm det}\,Y|^4\rangle=\frac{1}{6}
\frac{\Gamma(N+4)\Gamma(N+2)}{2^{2N}}\;.
\end{equation}

In order to evaluate
\begin{equation}
c_2=\frac{\langle |{\rm det}\,Z|^4(|a|^2+|{\bf w}^T Z^{-1}{\bf w}|^2)\rangle}{\langle
|{\rm
det}\,Y|^4\rangle}
,
\end{equation}
it is again profitable to use unitary invariance and turn ${\bf w}$ into
direction 1,
\begin{equation}
|{\bf w}^T Z^{-1}{\bf w}|^2=w^4|(Z^{-1})_{11}|^2\;.
\end{equation}
From
$\langle w^4\rangle=\frac{1}{4} N(N+1)$ and
$\langle |a|^2\rangle=1$ we obtain 
then the recursion relation
\begin{equation}
c_2(N)=\frac{4}{(N+1)(N+3)}+\frac{N}{N+3}c_2(N-1)\;,
\end{equation}
which is solved by
\begin{equation}
c_2(N)=\frac{2}{N+1}
\;.
\end{equation}
In order to reintroduce $\gamma$ we have to multiply
$c_m$ by $(\gamma N)^{m/2}$, and obtain to order $1/N$
\begin{equation}
c_2=2\gamma\;,\qquad c_4=4\gamma^2\;,
\end{equation}
as advertised above.

\section{Joint distribution of $B_1$ and $B_2$ for $\beta=2$}
\label{app:2}

For broken time-reversal symmetry, the distribution of
$B_1$ and $B_2$ has to be
calculated from the Laguerre ensemble (\ref{eq:laguerre}) with $\beta=2$.
Similarly as for preserved time-reversal symmetry, this
ensemble can be obtained from the eigenvalues of a matrix $W^\dagger W$.
The matrix $W$ is once more complex, but no longer symmetric (it is also not
Hermitian). It has the Gaussian distribution
\begin{equation}
P(W)\propto \exp(-2\gamma N{\rm tr}\,W^\dagger W)
,
\end{equation}
with measure
\begin{equation}
{\rm d}W=\prod_{i,j}{\rm d}\,{\rm Re}\,W_{ij}\,{\rm d}\,{\rm Im}\,W_{ij}
\,.
\end{equation}

It is instructive to calculate first $P(B_1)$,
because it will be instrumental in
the calculation of $P(B_1,B_2)$.
After averaging over the $u_i$'s, the characteristic function takes the form
\begin{eqnarray}
\chi(p)&=&\langle\exp(-ipB_1)\rangle
=\left\langle\frac{{\rm det}\,W^\dagger W}{{\rm det}\,(W^\dagger W+ip/N)}
\right\rangle
\;.
\end{eqnarray}
We express the determinant in the denominator as an integral over a 
complex vector ${\bf z}$.
Due to the invariance $W\to U W V$
of $P(W)$ for arbitrary unitary matrices $U$ and $V$,
we can turn ${\bf z}$ into direction 1 and write
\begin{equation}
W=\left(\begin{array}{cc}
a&
\begin{array}{cc}
x'&0^T
\end{array}
\\
\begin{array}{c}
x\\0
\end{array}
& Y
\end{array}
\right)
.
\end{equation}
Then
\begin{eqnarray}
&&P(B_1)\propto \int{\rm d}p\,{\rm d}z\,z^{2N-1}\,{\rm d}^2a\,
{\rm d}x\,x^{2N-3}
\,{\rm d}x'\,{x'}^{2N-3}
\nonumber\\
&&\quad{}\times
(|a|^2+d_2x^2x'^2)\exp[-(z^2+2\gamma N)(|a|^2+x^2)]
\nonumber\\
&&\quad{}\times\exp[i p(B_1-z^2/N)-2\gamma
Nx'^2]\;.
\end{eqnarray}
Selberg's integral \cite{Mehta} gives 
\begin{equation}
d_2\equiv \frac{\langle|{\rm det}\,Y|^2|(Y^{-1})_{11}|^2\rangle}{
\langle|{\rm det}\,Y|^2\rangle}=\frac{2\gamma N}{N-1}
.
\end{equation}
The integration over $p$ gives $\delta(z^2-NB_1)$ and allows to eliminate
$z$. The integration over $x'$ amounts to replacing
$x'^2=(N-1)/2\gamma N=d_2^{-1}$.
The final integrations are most easily carried out by concatenating $a$ to
${\bf x}$, giving an $N$-dimensional vector ${\bf y}$.
Then
\begin{eqnarray}
P(B_1)&\propto&\int{\rm d}y\,y^{2N+1}B_1^{N-1}
\exp[-N(B_1+2\gamma)y^2]\nonumber
\\
&\propto&B_1^{N-1}(B_1+2\gamma)^{-N-1}
,
\end{eqnarray}
which to order $1/N$ becomes
\begin{equation}
P(B_1)=\frac{2\gamma N}{B_1^2}\exp(-2\gamma N/B_1)
.
\label{eq:pbbeta2}
\end{equation}

The first steps in the calculation of the
joint distribution function of $B_1$ and $B_2$ are identical
to what was done in Appendix \ref{app:pab}, and result in the
characteristic function $\chi(p,q)$ in the form of
Eq.\ (\ref{eq:a5}), but with $\gamma$ replaced by $2\gamma$.
Due to the unitary invariance 
of the $W$ ensemble we can write
\begin{equation}
W=\left(\begin{array}{cc}
\begin{array}{cc}
a&x'\\
x&b
\end{array}
&
\begin{array}{cc}
0&0^T\\
y'&0^T
\end{array}
\\
\begin{array}{cc}
0&y\\
0&0
\end{array}
& Y
\end{array}
\right)
.
\end{equation}
One integrates now over $p$ and $q$ and obtains delta functions as in 
Eq.\ (\ref{eq:a14}).  This is followed by integration over $z$.
The calculation is then much simplified by recognizing that one can rescale
the remaining integration variables in such a way
(namely by introducing $a^2=\tilde a^2 B_1/B_2$, $x^2=\tilde x^2B_1/B_2$,
$y'^2=\tilde y'^2 x^{-2}B_2^{-1}$)
that
\begin{equation}
P(B_1,B_2)= B_2^{-3}\exp(-N B_1^2/B_2)f(B_1).
\end{equation}
It is not necessary to give here $f(B_1)$
as a lengthy multi-dimensional integral, since its functional form is
easily recovered from the relation
\begin{equation}
P(B_1)=\int{\rm d}B_2\,P(B_1,B_2)=N^{-2}B_1^{-4}f(B_1)
.
\end{equation}
We compare this with Eq.\ (\ref{eq:pbbeta2}) and arrive at
Eq.\ (\ref{eq:pbb2}).
The distribution of $B_2$ has the closed-form
expression
\begin{equation}
P(B_2)=\gamma^{-2}N^{-3}
G^{0,3}_{3,0}(\gamma^{-2}N^{-3}B_2|-\case{1}{2},-1,-2)
.
\end{equation}

\section{Joint distribution of $C_0$ and
$C_1$}
\label{app:a0a1}
We seek the joint distributions of the spectral moments
$C_0$ and $C_1$,  which determine
$\phi'$ and $I$ for $\beta=1$ and $n=m$ via Eq.\ (\ref{eq:phicd}).
We start with the characteristic function
\begin{equation}
\chi(p_0,p_1)=\langle\exp[i\,{\rm Re}\,(p_0 C_0+p_1
C_1)]\rangle
\;,
\end{equation}
where $p_0$ and $p_1$ are complex numbers, as are
the quantities $C_0$ and $C_1$
themselves.
Since $C_k=\sum_iu_i^2\tau_i^k$ we have to average over the $\tau_i$'s 
and the $u_i$'s.
Averaging over the $u_i$'s first, we obtain
\begin{eqnarray}
\chi(p_0,p_1)
&=&
\left\langle
\prod_i
\left(
1+\frac{|p_1\tau_i+p_0|^2}{N^2}
\right)^{-1/2}
\right\rangle
\;.
\end{eqnarray}
We regard again the rates $\mu_i=\tau_i^{-1}$ as the eigenvalues
of a matrix product $Y Y^\dagger$, where $Y$ will be specified below.
Then the product of square roots can be written as a ratio of determinants,
\begin{eqnarray}
&&\prod_i
\left(
1+\frac{|p_1\tau_i+p_0|^2}{N^2}
\right)^{-1/2}=
{\rm det}\,YY^T\nonumber\\
&&{}\times\det\left[(YY^T)^2\frac{N^2+|p_0|^2}{N^2}
+2\frac{{\rm Re}{p_0p_1^*}}{N^2}Y Y^T+\frac{|p_1|^2}{N^2}\right]^{-1/2}.
\nonumber\\
\end{eqnarray}

We will express the determinant in the denominator as a Gaussian integral over a {\em real}
$N$-dimensional vector ${\bf z}$. 
Hence it is convenient to choose $Y$ real as well, so that one can use orthogonal
invariance in order to turn ${\bf z}$ into direction 1.
Moreover, there is a representation of $Y$
which allows to incorporate the determinant
in the numerator into the probability measure: We take $Y$ as a rectangular $N
\times (N+3)$ matrix with random Gaussian 
variables, distributed according to
\begin{equation}
P(Y)\propto\exp(-\gamma N {\rm tr}\, Y Y^T)\;.
\end{equation}
The corresponding distribution of the eigenvalues $\mu_i$ of $Y Y^T$ 
is given in Ref.\ \cite{edelman}
and differs
from the Laguerre ensemble
(\ref{eq:laguerre}) by the
additional factor $\prod_i\mu_i={\rm det}\, Y Y^T$.
In this representation, 
\begin{eqnarray}
\chi&\propto& \int{\rm d}z\,z^{N-1}
\bigg\langle
\exp\left\{
-z^2(1+|p_0|^2/N^2)[(YY^T)^2]_{11}
\right\}
\nonumber\\
&&\quad{}\times \exp\left[
\frac{2{\rm Re}\,p_0p_1^*}{N^2}[YY^T]_{11}+\frac{|p_1|^2}{N^2}\right]
\bigg\rangle
\;,
\end{eqnarray}
where the average is now over $Y$.
Inverse Fourier transformation with respect to $p_0$ and $p_1$ results in
\begin{eqnarray}
&&P(C_0,C_1)\propto\left\langle \int{\rm d}z\,z^{N-5}
\frac{\exp[-z^2[(YY^T)^2]_{11}]}{[(YY^T)^2]_{11}-([YY^T]_{11})^2}
\right.
\nonumber\\
&&\left.{}\times
\exp\left[-\frac{|C_1|^2N^2}{4z^2}-\frac{N^2}{4z^2}
\frac{|C_0-[YY^T]_{11}C_1|^2}{[(YY^T)^2]_{11}-([YY^T]_{11})^2}
\right]\right\rangle\;.
\nonumber\\
\end{eqnarray}

The
orthogonal invariance of $YY^T$ allows us to parameterize $Y$ as
\begin{equation}
Y=\left(\begin{array}{cc}
\begin{array}{cc} a & v \end{array} 
&  0^T\\
\begin{array}{cc}
w & b \\
0 & y \\
0 & 0 
\end{array} 
&  Z
\end{array}
\right)
\;,
\end{equation}
with real numbers $v>0$, $w>0$, $y>0$, $a$, and $b$,
and an $(N-1)\times(N+1)$ dimensional 
matrix $Z$.
It is good to see that $Z$ drops out of the calculation,
because it does not appear in
\begin{mathletters}
\begin{eqnarray}
[YY^T]_{11}&=&a^2+v^2\;,\\
{}[(YY^T)^2]_{11}&=&(a^2+v^2)^2+(aw+vb)^2+v^2y^2\;.
\end{eqnarray}
\end{mathletters}
We replace $b=b'-aw/v$ and introduce $z'=zyv$.
The integral over $z'$ can be written in the saddle-point form
$\int{\rm d}z'\, {z'}^{N}e^{-z'^2}f(z')\propto f(\sqrt{N/2})$
for large $N$.
The resulting expression
varies with respect to the remaining variables on the scales
\begin{equation}
N^3 a^2 \simeq N^2 {b'}^2\simeq N^2 v^2
\simeq N y^2\simeq w^2={\cal O}(\gamma^{-1})\;.
\end{equation}
We use the given orders of
magnitude to eliminate terms of order $N^{-1}$,
but  keep the residual correlations
${\rm Re}\,C_0 C_1^*/\gamma N={\cal O}(N^{-1/2})$.
The joint distribution function of $C_0$ and $C_1$ is then
\begin{eqnarray}
&&P(C_0,C_1)\propto
\int
{\rm d}a\,{\rm d}b'\,{\rm d}v\, v^3\,{\rm d}w\, w^{N-2}\,{\rm d}y\,y
\exp[-\gamma N y^2]
\nonumber\\
&&{}\times\,
\exp\left[-\gamma N w^2\left(1+\frac{a^2}{v^{2}}\right)
-\frac{N}{2y^2}(v^2+y^2+{b'}^2)\right]
\nonumber\\
&&{}\times\,
\exp\left[
 N v^2{\rm Re}\,C_0 C_1^*
-\frac{N v^2y^2|C_1|^2}{2}-\frac{N|C_0|^2}{2}
\right]
\;.
\end{eqnarray}
Now we can integrate over $a$, $b'$, $w$,  
and $v$, and arrive at
\begin{eqnarray}
&&P(C_0,C_1)\propto
\int
{\rm d}y
\frac{\exp[-\gamma N y^2-N|C_0|^2/2]}{
\left( y^{-2}
+y^2|C_1|^2+2{\rm Re}\,C_0 C_1^*\right)^{5/2}}
\;.
\nonumber\\
\end{eqnarray}
The final result (\ref{eq:pcdnn}) is obtained by substituting $s=\gamma Ny^2$.

\section{Distribution of $A_1$ for $\beta=1$}
\label{appa1}

In the large-$N$ limit
the joint distribution function $P(A_0,A_1)=P(A_0)P(A_1)$ factorizes
as explained  in
Section \ref{sec:qual}. The distribution of $A_0$
is given in Eq.\ (\ref{eq:pa0}).
It remains to calculate the distribution of
$A_1=\sum_i\tau_iu_iv_i$.
The $u_i$'s and $v_i$'s are independent Gaussian random numbers.
Averaging over them,
we obtain the characteristic function
\begin{eqnarray}
\chi(p)&=&\langle\exp[i\,{\rm Re}\,(p A_1)]\rangle
\nonumber \\
&=&
\left\langle
\prod_i
\left(
1+\frac{|p\tau_i|^2}{4N^2}
\right)^{-1}
\right\rangle
\nonumber \\
&=&
\left\langle
\frac{{\rm det}\,(W^\dagger W)^2}{{\rm det}[(W^\dagger W)^2+|p|^2/4N]}
\right\rangle
\;,
\end{eqnarray}
where $p$ is a complex number.
The Laguerre ensemble is again represented 
as the eigenvalues of the matrix product $W^\dagger W$, where $W$ is the
complex symmetric matrix with distribution (\ref{eq:pw1}).
Following the route of
Appendix \ref{app:pab} we represent the determinant in
the denominator by a Gaussian integral over a complex vector ${\bf z}$
and choose a basis in which $W$ is of the form (\ref{eq:parw}).
The characteristic function is then obtained as the following multi-dimensional 
integral,
\begin{eqnarray}
&&\chi(p)=\int{\rm d}{\bf x}\,{\rm d}{\bf y}\,{\rm d}{\bf z}
\,{\rm d}^2a\,{\rm d}^2b\,{\rm d}Y
\nonumber\\
&&\quad{}\times |{\rm det}\,Y|^4|a[b-y^2(Y^{-1})_{11}]^2-x^2|^4
\exp\left(-\frac{|zp|^2}{4N^2}\right)
\nonumber\\
&&\quad{}\times 
\exp\left[-z^2\left((|a|^2+x^2)^2+x^2|a+b^*|^2+x^2y^2
\right)\right]
\nonumber\\
&&\quad{}\times
\exp\left[-\gamma N(|a|^2+|b|^2+2x^2+2y^2+{\rm tr}\,Y^\dagger Y)\right]
.
\nonumber\\
\end{eqnarray}

Let us briefly describe in which order the integrations are performed most
conveniently.
Fourier transformation with respect to $p$ converts the characteristic
function back into the distribution function $P(A_1)$. This step
gives rise to a factor
$z^{-2}\exp(-|A_1|^2N^2 z^{-2})$. We can also integrate over ${\bf y}$, 
which results in a factor $\exp[-2\gamma N/(xz)^2]$. 
We introduce new variables by the substitutions
$b=\tilde b-a^*$, ${\bf x}={\bf v}/z$, $a=a'/z$.
After these transformations one
succeeds  to integrate over $b'$, ${\bf z}$, and $a'$. The remaining integral
over $v=|{\bf v}|$ is of the form
\begin{eqnarray}
&&P(A_1)\propto |A_1|^{-5}\int\,{\rm d}v\,v^{-5}e^{-2/v}
\nonumber\\
&&\quad{}\times\bigg[
\pi[8|A_1|^2(2+4v+v^2)+3v^2(16+16v+3v^2)]
\nonumber\\
&&\quad{}
-\frac{2|A_1|v}{(|A_1|^2+v^2)^4}[
|A_1|^4v^4(288+304v-25v^2)
\nonumber\\
&&\quad{}\,
+|A_1|^6v^2(192+176v-17v^2)
+8|A_1|^8(6+4v-v^2)
\nonumber\\
&&\quad{}\,
+3v^8(16+16v+3v^2)
+|A_1|^2v^6(192+208v+41v^2)]
\nonumber\\
&&\quad{}
-[16|A_1|^2(2+4v+v^2)+6v^2(16+16v+3v^2)]
\nonumber\\
&&\qquad{}\times \arctan\left(\frac{v}{|A_1|}\right)
\bigg].
\end{eqnarray}
The more compact form (\ref{eq:pa1t}) is the result of the replacement $v=2/s$,
followed by a number of partial integrations.

\end{document}